\documentclass[conference]{IEEEtran}

\ifCLASSINFOpdf
\usepackage{fancyhdr}
\usepackage{graphicx}
\usepackage{textcomp}
\usepackage{fixltx2e}
\usepackage{listings}
\usepackage{enumerate}
\usepackage[shortlabels]{enumitem}
\usepackage{xcolor}
\usepackage{multirow}
\usepackage{hhline}
\usepackage{amsmath}
\usepackage[utf8]{inputenc}
\usepackage{fourier} 
\usepackage{array}
\usepackage{makecell}
\usepackage{url}
\usepackage{dblfloatfix}

\else
\fi

\makeatletter

\renewcommand{\thispagestyle}[1]{}
\date{}

\let\old@ps@headings\ps@headings
\let\old@ps@IEEEtitlepagestyle\ps@IEEEtitlepagestyle
\def\confheader#1{%
    \def\ps@headings{%
        \old@ps@headings%
        \def\@oddhead{\strut\hfill#1\hfill\strut}%
        \def\@evenhead{\strut\hfill#1\hfill\strut}%
    }%
    \def\ps@IEEEtitlepagestyle{%
        \old@ps@IEEEtitlepagestyle%
        \def\@oddhead{\strut\hfill#1\hfill\strut}%
        \def\@evenhead{\strut\hfill#1\hfill\strut}%
    }%
    \ps@headings%
}
\makeatother

\title{Hybrid NER System for Multi-Source Offer Feeds\textsuperscript{\fontsize{12}{20} \selectfont 1}}
\begin{document}

\newcolumntype{C}[1]{>{\centering\let\newline\\\arraybackslash\hspace{0pt}}m{#1}}

\author{
  \hspace*{-0.5cm}\begin{tabular}{C{6cm}C{6cm}C{6cm}}
    Anusha Holla \newline \textit{Samsung Research Institute} \newline \textit{Bangalore, India} \newline anusha.holla@gmail.com
 &
    Bharat Gaind \newline \textit{Samsung Research Institute} \newline \textit{Bangalore, India} \newline bharatgaind234@gmail.com
 &
   Vikas Reddy Katta \newline \textit{Samsung Research Institute} \newline \textit{Bangalore, India} \newline vikasreddy.k11@gmail.com
  \end{tabular} \\ \\
    	
  \begin{tabular}{C{6cm}C{6cm}}
    Abhishek Kundu \newline \textit{Samsung Research Institute} \newline \textit{Bangalore, India} \newline kundu.abhishek@gmail.com
 &
   S Kamalesh \newline \textit{Samsung Research Institute} \newline \textit{Bangalore, India} \newline kamalsam.21@gmail.com
	\end{tabular}
}

%

\maketitle

\begin{abstract}
Data available across the web is largely unstructured. Offers published by multiple sources like banks, digital wallets, merchants, etc., are one of the most accessed advertising data in today’s world. This data gets accessed by millions of people on a daily basis and is easily interpreted by humans, but since it is largely unstructured and diverse, using an algorithmic way to extract meaningful information out of these offers is hard. Identifying the essential offer entities (for instance, its amount, the product on which the offer is applicable, the merchant providing the offer, etc.) from these offers plays a vital role in targeting the right customers to improve sales. This work presents and evaluates various existing Named Entity Recognizer (NER) models which can identify the required entities from offer feeds. We also propose a novel Hybrid NER model constructed by two-level stacking of Conditional Random Field, Bidirectional LSTM and Spacy models at the first level and an SVM classifier at the second. The proposed hybrid model has been tested on offer feeds collected from multiple sources and has shown better performance in the offer domain when compared to the existing models.

\end{abstract}

\IEEEpeerreviewmaketitle

\section{INTRODUCTION}
Offers are one of the major sources of unstructured data in the marketing domain. They are also one of the most consumed datasets. Every single day, millions of customers read offer statements and extract meaning out of them, which they use for improving the profitability of their shopping experience. It would be highly beneficial for the industry to use this wealth of data to enhance existing customer shopping experience. If offers can be converted to a machine-readable format, algorithms could be developed to target the right customers, which can prove vital in improving sales. The motivation is to analyze marketing offers based on information extraction, in an industrial setting. One use-case where extracting the constituent entities/attributes of offers could be important is an organization/business trying to understand the offers that are being offered by their competitors in the market. The solutions proposed in this paper could be utilized by a third-party business to create a portal where marketing offers of these competitors could be compared, using which the buisness can provide a better offer to their customers and thus, improving sales. Another use-case could be to filter all the unnecessary offers received by the user (as SMS messages on his phone) to give him/her personalized offers and avoid clutter. Yet another use-case could be a continuation of the work done by Ujwal et al. \cite{ujwal2017classification}, which proposes a method to scrape offers from offer-aggregator websites. The Hybrid Model we propose could be used to extract meaningful entities from these scraped offers. All this is only possible if the essential elements that make up the offers are correctly understood. 

However, there are multiple challenges in doing this. One of these challenges is the problem of data variety. Offers come from numerous sources in various formats - all in natural language. It is difficult to convert these offers to a machine-readable format (like JSON). Also, the structure of the offers from a source is prone to vary. In this paper, we try to address these challenges and enhance the prediction accuracy by proposing a novel Hybrid Named Entity Recognition (NER) system, constructed by two-level stacking of Conditional Random Field (CRF), Bidirectional LSTM and spaCy \cite{spacy} models in the first level and a Support Vector Machine (SVM) classifier in the second. These models have been implemented using some very popular Natural Language Processing (NLP) and Machine Learning (ML) libraries, such as Stanford NER \cite{StanfordNER}, Keras \cite{Keras}, spaCy and scikit-learn \cite{scikit}. We also evaluate and compare the independent NER models (the ones used at the first level: CRF, BLSTM, spaCy) and the Hybrid Model by training them on four known sources and subsequently testing them on an unknown fifth one. It is found that the proposed Hybrid Model has a significantly higher accuracy when compared to the other models. Therefore, it can be used to efficiently extract various important entities in offer feeds.

\renewcommand{\arraystretch}{2}
\begin{table*}[]
\centering
\caption{Data Sources}
\label{sources}
\begin{tabular}{|l|l|l|l|l|l|}
\hline
\textbf{Datasets} & \textbf{\thead{Dataset\\Source}} & \textbf{\thead{Source Url}}                                                & \textbf{\thead{Number of\\Offers  Scraped}} & \textbf{\thead{Number of\\Templates made}} & \textbf{\thead{Number of Offers\\after bloating}} \\ \hline
$D_1$                & Axis Bank               & https://www.axisbank.com/grab-deals/online-offers                  & \makecell{91}                                & \makecell{35}                                & \makecell{651}                                      \\ \hline
$D_2$                & ICICI Bank              & \makecell{https://www.icicibank.com/Personal-Banking/offers\\/offer-index.page} & \makecell{95}                                & \makecell{27}                                & \makecell{864}                                      \\ \hline
$D_3$                & HDFC Bank               & \makecell{https://offers.smartbuy.hdfcbank.com/list\_offer\\/credit\_card/2   } & \makecell{42}                                & \makecell{33}                                & \makecell{761}                                      \\ \hline
$D_4$                & Grabon                  & https://www.grabon.in/paytm-coupons/                               & \makecell{148}                               & \makecell{34}                                & \makecell{891}                                      \\ \hline
$D_5$                & SBI Bank                & https://www.sbicard.com/en/personal/offers.page                    & \makecell{14}                                & \makecell{10}                               & \makecell{57}                                       \\ \hline
\end{tabular}
\end{table*}
\section{LITERATURE REVIEW}
Named Entity Recognition is a subtask of information extraction that seeks to locate and classify named entities in text into pre-defined categories \cite{wikiNER}. There are a number of algorithms that can be used for Named Entity Recognition. Various Named Entity Recognition systems have been developed in the last two decades. But, there has not been a significant effort to analyze the complex marketing offers, which is a very important domain (as explained in the previous section). In the effort of building NERs in the offer domain, we have drawn inspiration from various previous works/literature.

Initially, statistical methods were commonly applied to build Named Entity Recognizers \cite{Lafferty:2001:CRF:645530.655813}. Recently, neural architectures have gained popularity for Named Entity Recognition. The work of Zhiheng et al. \cite{BiLSTM} discusses the Bidirectional LSTM for sequential Tagging. The work of Shriberg et al. \cite{citeulike} and  Lafferty et al. \cite{Lafferty:2001:CRF:645530.655813} has shown that CRFs can produce higher tagging accuracy. Comparisons made by R.Jiang et al. \cite{spacyJiang} showed that spaCy performed best, next to Stanford NER. Another method is Stacking, which allows blended intelligence from many different approaches to be combined into one superior result. Stacked generalization was introduced by Wolpert \cite{Wolpert92stackedgeneralization}. We take inspiration from various concepts/works described above to build our proposed Hybrid system, which shows significantly better results than any of the existing/popular NER systems (also evaluated in this paper), in the marketing offers domain.

\section{DATASET}
The offer-data is collected by scraping offers from five different sources. Four of these sources are banks, and the fifth is an offer-aggregator website. The offers contained in each of these sources are very diverse and different in structure from one another. Each offer contains some entities/attributes that constitute the offer. We call each such entity a \textbf{tag}. The following is the list of tags in an offer that we are interested in extracting:

\begin{itemize}
    \item \emph{OAMT} - Offer amount
    \item \emph{OTYPE} - Offer Type (discount, cashback, voucher)
    \item \emph{MIN\_AMT} - Minimum purchase amount above which offer is valid
    \item \emph{MAX\_AMT} - Maximum offer amount
    \item \emph{PRD} - Product on which the offer is valid
    \item \emph{MERCH} - Name of the Merchant offering the Offer
    \item \emph{O} - Any token we're not interested in extracting as an offer-entity, should be tagged as Other (O).
\end{itemize}

Since the number of offers obtainable from these sources is limited in number and not enough to train an NER model, we use \textbf{offer-templates} (generic structures that the maker of the offer follows, while creating the offer) to generate a large number of offers. For example, the offer, \emph{``Get 20\% off on pizzas at Dominos"} follows the generic offer-template, \emph{``Get OAMT OTYPE on PRD at MERCH"} (where \emph{OAMT}, \emph{OTYPE}, etc. are tags). We now convert the scraped offers from each source into its corresponding set of offer-templates. Five different labeled datasets (containing a large number of offers) are created corresponding to each of these five sets of offer-templates, after bloating their (offer-templates') constituent tags randomly with appropriate values. Finally, we tokenize all these datasets. To tokenize the input uniformly for all our NER models, we use the spaCy tokenizer. The resultant labeled datasets are called the \textbf{tokenized datasets}, which will be subsequently used for supervised learning. For simplicity, we refer to them as \textbf{\emph{D\textsubscript{i} (i=1,2,..5)}}. Four of these datasets ($D_{1}$, $D_{2}$, $D_{3}$, $D_{4}$) are used for training and the fifth one for testing ($D_{5}$ or \textbf{\emph{D\textsubscript{test}}}). The details of these datasets are shown in Table \ref{sources}.

\section{SYSTEM ARCHITECTURE}
In this paper, we use three independent models for the purpose of Named Entity Recognition (NER): CRF Model, BLSTM Model, and spaCy Model. Then, we use an SVM Classifier to combine these models and propose a Hybrid Model.

\subsection{CRF Model}

Conditional Random Field (CRF) is a probabilistic sequence model, mainly used for NER. It is a framework for building probabilistic models to segment and label sequential data. It is preferred because they offer a huge advantage by relaxing the independence assumptions made by models like HMMs (Hidden Markov Models) and stochastic grammars \cite{Lafferty:2001:CRF:645530.655813}.

In this paper, we use Stanford NER to implement the CRF classifier, which has a Java-based implementation of the same. It expects its input (a tokenized dataset) as pairs of tab-separated tokens (words) and tags, in separate lines, where each offer-message is separated by two new lines. The following features are set to \emph{true} in Stanford NER while training the CRF model:
\begin{itemize}
\item usePrev
\item useNext
\item useTags
\item useWordPairs
\item usePrevSequences
\item useNextsequences
\item useLemmas
\item useLemmaAsWord
\item normalizeTerms
\item normalizeTimex
\item usePosition
\item useBeginSent
\end{itemize}
The output generated by this model is the probability of each tag for every token. 

Now, there could be instances in the future, where offers are coming from a new unknown source. Also, the structure of offers coming from a particular source is prone to vary. Hence, there is a need for a system, which is agnostic to the source of an offer. So, it is better to combine all the tokenized training datasets ($D_{1}$, $D_{2}$, $D_{3}$, $D_{4}$) into a single \textbf{combined dataset \emph{D\textsubscript{comb}}}, so that the final dataset used for training contains as many diverse offer-templates as possible. To further justify the need of a combined dataset, we experimented by training various CRF models on individual datasets ($D_{1}$, $D_{2}$, $D_{3}$, $D_{4}$) and another model on the combined dataset. It was found (see results in Section V) that the accuracy was higher for the combined dataset model, compared to the individual dataset models. \emph{D\textsubscript{comb}} is further divided in two equal sets : \textbf{\emph{D\textsubscript{comb1}}} and \textbf{\emph{D\textsubscript{comb2}}}. \emph{D\textsubscript{comb1}} is used to train the three independent models (CRF, BLSTM and spaCy) and \emph{D\textsubscript{comb2}} is used to train the Hybrid model. The CRF model trained using the dataset $D_{comb1}$ is referred to as \textbf{\emph{M\textsubscript{CRF}}}.

\subsection{BLSTM Model}

In the last few years, Recurrent Neural Networks (RNNs) have shown significant results in a variety of tasks like speech recognition, language modeling, translation, and image captioning. The idea of RNNs is that they use previous information while predicting the tag for the current token (word). Consider the offer, \emph{``Shop at Lifestyle and get flat 20\% off on apparels"} and the offer, \emph{``Get instant 20\% off on Lifestyle"}. In the first example, the token followed by \emph{``on"} (the last token of the sentence) should be tagged as \emph{PRD}, whereas in the second example, the token followed by \emph{``on"} should be tagged as \emph{MERCH}. To predict what comes after \emph{``on"}, we need a history of what has already been seen in the sentence. RNNs don't seem to be able to learn long-term dependencies \cite{lstmBengio}, which is why Long Short Term Memory (LSTM) is needed. In the first example, the information that \emph{MERCH} was already seen at the beginning of the sentence can be used by an LSTM model to predict what comes after \emph{``on"} (\emph{PRD} in this case). Also, since we need to consider both the left and the right side long-term dependencies of a token while predicting its tag accurately, we need to use Bi-directional LSTM (BLSTM) \cite{blstmgraves} for the purpose of NER.

The BLSTM model is implemented using Keras. It is trained using the dataset $D_{comb1}$ (as explained in the previous section). The input to the model is a list, where each element is itself a list of pairs of tokens and tags of an offer-message. Each of the tokens in an offer-message is converted to one-hot encoding and GloVe embedding \cite{glove} is applied to get a 300-dimensional vector, corresponding to every token. Each offer-message is padded with zeroes to make the size of all the offer-messages equal. The output from the hidden states is a 64-dimensional vector which is applied over softmax activation function to get a \emph{7-dimensional vector} (because the number of tags is 7). This vector represents the probability scores of tags for every token. The BLSTM model thus built is represented as \textbf{\emph{M\textsubscript{BLSTM}}}.

\subsection{spaCy Model}

spaCy is an open-source software library for advanced Natural Language Processing, written in Python and Cython. Ridong Jiang et al. \cite{spacyJiang} showed that spaCy performed best, next to Stanford NER. 

The expected input for spaCy is a list, where every element is itself a list of the offer-message sentence, the start and end index in that sentence of the token that corresponds to a tag, and finally, the tag itself. For training, we used the default English model in spaCy. This model is also trained using the tokenized dataset $D_{comb1}$. The tokens from $D_{comb1}$ are fed into spaCy's EntityRecognizer. It generates docs (a sequence of tokens) for each offer-message, which when fed into the GoldParse, along with the tag offsets (a list of tag locations in the offer-message), produces gold-standard tokens. These tokens and their associated tags are then fed to spaCy's EntityTagger to train the model. The model is updated (retrained) for every offer-message. The output of this model is the tag associated with each token, whereas the list of probabilities associated with the tokens is not given. The model built from spaCy is represented as \textbf{\emph{M\textsubscript{spaCy}}}.

\subsection{The Hybrid Model}
In each of the models explained above, we are relying on a single model for entity recognition. But, diversification of models provides a more robust prediction. Hence, ensembling is used. Ensembling is a technique of combining the individual predictions of multiple models to give superior results. The resulting model is often much more accurate than the constituent individual classifiers \cite{EnsembleClassifiers1}, \cite{EnsembleClassifiers2}.
\\\
\begin{figure}[b!]
\begin{center}
\includegraphics[height=8cm]{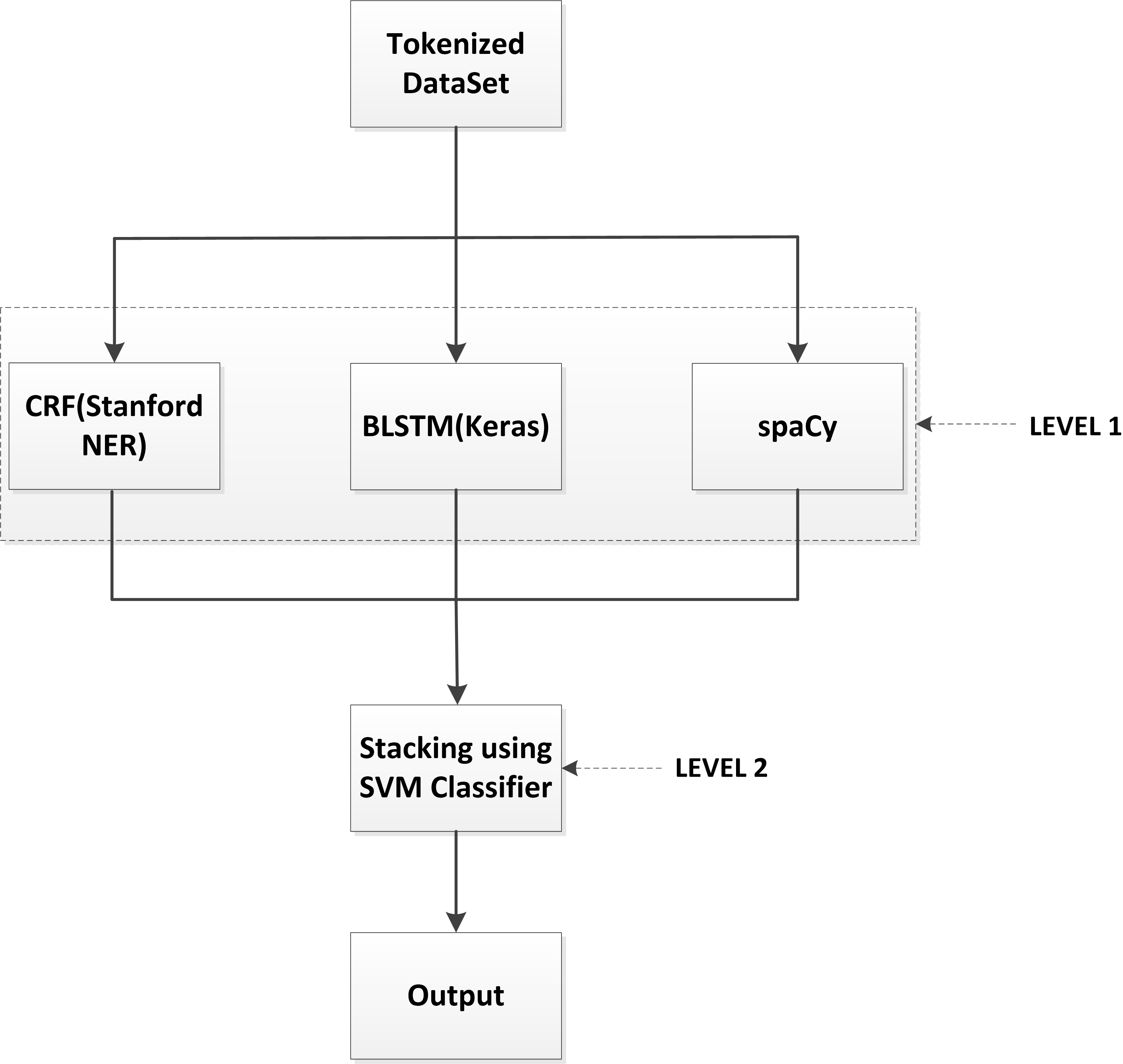}
\end{center}
\caption{System Architecture Diagram}
\label{fig:architecure}
\end{figure}

\par There are three main methods of ensembling: \emph{Bagging}, \emph{Boosting} and \emph{Stacking}. \emph{Bagging} (stands for Bootstrap Aggregation) improves the classification by combining classifications of randomly generated training sets \cite{Breiman1996}. It is aimed to decrease variance. In the case of \emph{Boosting}, the results of previous classifier's misclassified data are used to train the next classifier. All the classifiers are aggregated using majority voting. It is aimed to decrease bias. In \emph{Stacking}, we use a pool of base classifiers, and then use another classifier to combine the predictions, with the aim of reducing the generalization error. Since our application requires to reduce both the variance and bias, we make use of stacking. The stacked model will be able to discern where each model performs well and where it performs poorly.

The Hybrid Model, we propose, is constructed using two-level stacking. Three models are used at the first level: $M_{CRF}$, $M_{BLSTM}$ and $M_{spaCy}$ (as trained in the previous sections). A Linear SVM classifier is used at the second level. It is a standard method for large-scale classification tasks and is preferred because it is one of the best multi-class text classifiers. This classifier is implemented using scikit-learn's SVMClassifier, with Hinge Loss function. The two levels of the Hybrid model are depicted in Fig. \ref {fig:architecure}.

The following steps are used for training the Hybrid Model:
\begin{itemize}
	\item First, we feed the dataset $D_{comb2}$ as input to  \emph{M\textsubscript{CRF}}, \emph{M\textsubscript{BLSTM}}, \emph{M\textsubscript{spaCy}}.
	\item For every token, the output of \emph{M\textsubscript{CRF}} (a \emph{7-dimensional vector} of the probabilities of all 7 tags for every token), \emph{M\textsubscript{BLSTM}} (another \emph{7-dimensional vector} of the probabilities of all 7 tags for every token) and \emph{M\textsubscript{spaCy}} (an integer in the range [0, 5] depicting the tag predicted for a token) is merged to form a \emph{15-dimensional vector}.
	\item A list (\emph{\textbf{l\textsubscript{X}}}) of such \emph{15-dimensional vectors} (with each vector representing a token), created by merging all the tokens in all the offers in $D_{comb2}$, is fed as input to train the SVM classifier. Another list (\emph{\textbf{l\textsubscript{Y}}}) containing the correct tags (already present in the dataset) for each of the tokens is also fed as input to the classifier. For example, if there are 100 offers, and each offer has an average of 10 tokens, $l_X$ will have 1000 \emph{15-dimensional vectors}, whereas $l_Y$ will contain 1000 correct tags, corresponding to each of the tokens.
\end{itemize}

The output of the model is the tag associated with each token (word) of an offer-message. The Hybrid model, thus formed, is represented as \textbf{\emph{M\textsubscript{Hybrid}}}.

\section{RESULTS AND DISCUSSION} \label{results}

In this section, we test the various models we trained in the previous sections: $M_{CRF}$, $M_{BLSTM}$, $M_{spaCy}$ and $M_{Hybrid}$, using the metric F1 score/F Measure. But before that, we define the various metrics, needed to evaluate the F1 score of our models:

\begin{itemize}
\item True Positive (TP): The token is correctly classified as one of the six tags: \emph{OAMT}, \emph{OTYPE}, \emph{MIN\_AMT}, \emph{MAX\_AMT}, \emph{PRD} and \emph{MERCH}.

\item True Negative (TN): The token is correctly classified as the tag \emph{O} (which is not a tag we're interested in extracting).

\item False Positive (FP): The token is misclassified as one of the six tags: \emph{OAMT}, \emph{OTYPE}, \emph{MIN\_AMT}, \emph{MAX\_AMT}, \emph{PRD} and \emph{MERCH}.

\item False Negative (FN): The token is misclassified as the tag \emph{O}.
\end{itemize}

The precision, recall and finally the F1 score are calculated using the following formulas:

\begin{equation}
Recall = \frac{TP}{TP + FN}
\end{equation}

\begin{equation}
Precision = \frac{TP}{TP + FP}
\end{equation}

\begin{equation}
F1\ score = \frac{2*Precision*Recall}{Precision + Recall}
\end{equation}
\thinspace
\thinspace
\begin{table}[h]
\centering
\caption{Comparison of various CRF Models}
\label{crf_models}
\begin{tabular}{|c|c|}
\hline
\textbf{CRF Models} & \textbf{F1 score} \\ \hline
$M_{CRF1}$               & 0.5125            \\ \hline
$M_{CRF2}$               & 0.5497            \\ \hline
$M_{CRF3}$               & 0.4618            \\ \hline
$M_{CRF4}$               & 0.4044            \\ \hline
\textbf{\emph{M\textsubscript{CRF}}}                & \textbf{0.6130}            \\ \hline
\end{tabular}
\end{table}

Before proceeding with the testing of various models trained, we first prove that a combined dataset model ($D_{comb1}$) will give better accuracy than the models trained on individual datasets: $D_1$, $D_{2}$, $D_{3}$, $D_{4}$ (as explained in Section IVA). For this, we train four CRF models, ${M_{CRF1}}$, ${M_{CRF2}}$, ${M_{CRF3}}$, ${M_{CRF4}}$, corresponding to the datasets, $D_1$, $D_{2}$, $D_{3}$, $D_{4}$ and use the already trained CRF model, ${M_{CRF}}$, corresponding to the dataset, $D_{comb1}$ (trained in section IVA). We tested all these five models on $D_{test}$, as shown in Table \ref{crf_models}. It can be seen that the accuracy of ${M_{CRF}}$ is higher than the accuracy of the models trained on the individual datasets, which further justifies the need to diversify the datasets by combining them.

\renewcommand{\arraystretch}{2}

\begin{table}[h]
\centering
\caption{Overall F1 scores of the various models}
\label{table:composite_and_hybrid}
\begin{tabular}{|c|c|}
\hline
\textbf{Models} & \textbf{F1 score} \\ \hline
$M_{CRF}$            & 0.6130            \\ \hline
$M_{BLSTM}$          & 0.7761            \\ \hline
$M_{spaCy}$          & 0.6870            \\ \hline
\textbf{\emph{M\textsubscript{Hybrid}}}        & \textbf{0.8156}            \\ \hline
\end{tabular}
\end{table}

\begin{table}[h]
\centering
\caption{Tag Wise F1 Scores of the various models}
\label{table:individual_tags}
\begin{tabular}{|l|l|l|l|l|}
\hline
         & \textbf{\emph{M\textsubscript{CRF}}} & \textbf{\emph{M\textsubscript{BLSTM}}} & \textbf{\emph{M\textsubscript{spaCy}}} & \textbf{\emph{M\textsubscript{Hybrid}}} \\ \hline
OAMT     & 0.7742        & 0.8110          & 0.6987          & 0.8366  \\ \hline
OTYPE    & 0.6992        & 0.8571          & 0.7717         & 0.9714  \\ \hline
MIN\_AMT & 0.4545        & 0.7397          & 0.6857         & 0.8750  \\ \hline
MAX\_AMT & 0.1739        & 0.5945          & 0.0            & 0.7050  \\ \hline
PRD      & 0.4706        & 0.8750 & 0.7407         & 0.8478           \\ \hline
MERCH    & 0.5714        & 0.6560 & 0.5870         & 0.6458           \\ \hline
\end{tabular}
\end{table}

Now, we test the models, $M_{CRF}$, $M_{BLSTM}$, $M_{spaCy}$ and $M_{Hybrid}$ on $D_{test}$. The overall F1 scores (calculated using the total TPs, FNs and FPs across all tags) for all models is shown in Table \ref{table:composite_and_hybrid}. Also, the F1 scores of all 6 tags for each of the models is shown in Table \ref{table:individual_tags}.

The proposed Hybrid Model was tested on the same dataset as the rest of the models, and as we can see, the F1 score of the last row in Table \ref{table:composite_and_hybrid} is significantly higher compared to the other models. The Hybrid Model is 3.95\% more accurate than the BLSTM Model, which is the most accurate among the three independent models (CRF, BLSTM, spaCy). The reason for this is that while training, the hybrid model assigns different weights to different models, based on their performances on the various tags. In other words, an informed decision is made and accordingly more weights are assigned to the better performing models for a particular tag. The better performance of the proposed model is also evident from the tag wise F1 scores reported in Table \ref{table:individual_tags}, where its accuracy is higher on almost all the tags when compared to the other models. Another important point to be observed here is that since the dataset $D_{test}$ is completely unknown to the hybrid model, it simulates the case when the offer-structure has been changed in a known-source (which was used to train the model). Therefore, the good performance of the hybrid model indicates/implies that the problem of structure change of an offer-source has been addressed.

\section{CONCLUSION}
 
In this paper, we evaluate the various existing/popular NER models (CRF, BLSTM, spaCy) to analyze marketing offers, in an industrial setting. We also propose a Hybrid model, constructed by two-level stacking. Amongst all the models, the Hybrid Model gives the best results, when tested on an unknown source. We also try to solve the problem of data variety and structure-change, using this model. This work can be further extended by training on more than four sources, so as to get better accuracies. Furthermore, apart from the marketing offer domain, the proposed Hybrid Model can be extended to other domains of interest as well.

\bibliography{Master}
\bibliographystyle{ieeetr}

\end{document}